\documentclass[
 reprint,amsmath,amssymb,superscriptaddress
]{revtex4-2}

\usepackage{graphicx}
\usepackage{dcolumn}
\usepackage{bm}

\usepackage{amsmath}

\DeclareSymbolFont{timesroman}{T1}{ptm}{m}{it}
\DeclareMathSymbol{\mathQ}{\mathalpha}{timesroman}{`Q}

\begin{document}

\title{Thermally Activated Snap-through Transitions Controlled by Tunable Metastability} 

\author{Renjie Zhao}
\email[E-mail: ]{renjiephys@gmail.com}
\affiliation{Department of Physics, Nanjing Normal University, and Key Laboratory of Numerical Simulation for Large Scale Complex Systems, \\Ministry of Education, Nanjing 210023, China
}
\author{Yiquan Zhang}
\email[E-mail: ]{zhangyiquan@njnu.edu.cn}
\affiliation{Department of Physics, Nanjing Normal University, and Key Laboratory of Numerical Simulation for Large Scale Complex Systems, \\Ministry of Education, Nanjing 210023, China
}
\author{Chenglin Luo}
\affiliation{Department of Physics, Nanjing Normal University, and Key Laboratory of Numerical Simulation for Large Scale Complex Systems, \\Ministry of Education, Nanjing 210023, China
}
\author{Yihang Wang}
\email[E-mail: ]{yxw2626@case.edu}
\affiliation{Department of Chemistry, Case Western Reserve University, Cleveland, OH 44106, United States}

\date{\today}

\begin{abstract}

The effects of thermal fluctuations on the morphology of two-dimensional materials are hard to harness. We propose that a geometrically constrained graphene nanoribbon (GNR) can exhibit thermally activated snap-through transitions with a predictable and controllable transition rate constant. The energetics and kinetics of such transitions can be fully captured by combining enhanced sampling methods and generalized transition state theory. Using well-tempered metadynamics, we determine the free energy landscape and a pair of asymmetric transition pathways of the GNR system. Notably, generalized transition state theory accurately captures how the transition rate constant responds to temperature and the tunable free energy landscape of our system. This work offers a theoretical framework for elastic metastability, introduces rare event methods into thermalized nanomechanical systems, and provides potential applications in designing nanoscale thermal switches and thermal actuators.

\end{abstract}

\maketitle


Thermally activated transitions underlie a great variety of processes in physics, chemistry, and biology, with examples ranging from magnetization reversal \cite{myers2002thermally,brown2001langevin} and chemical reactions \cite{hanggi1990reaction,peters2017reaction} to crystal nucleation~\cite{wolde1997enhancement} and protein folding \cite{s1994does}. A common characteristic of these systems is the presence of a free energy landscape with metastable states, between which transitions are driven by thermal fluctuations. We propose that elastic structures associated with snap-through transitions~\cite{gomez2017critical,radisson2023elastic,wang2024transient}, when miniaturized to the nanoscale, constitute an underexplored class of systems where metastable free energy landscapes can emerge and give rise to thermally activated morphological transitions.

The morphological snap-through transition occurs when a bistable elastic structure rapidly switches from one equilibrium state to the other. These transitions allow the generation of fast movements through the release of stored elastic energy and are manifested in various biological and engineered systems, including the Venus flytrap \cite{forterre2005venus}, hummingbird beaks \cite{smith2011elastic}, mechanical metamaterials \cite{silverberg2014using}, and nanoelectromechanical switches \cite{loh2012nanoelectromechanical}. In macroscopic systems, a snap-through transition is typically realized either by applying an external load that overcomes the energy barrier, or by imposing boundary actuation that destabilizes one of the equilibrium states \cite{gomez2018ghosts}. However, due to the presence of thermal fluctuations in microscopic systems, such external interventions may not be required for a snap-through transition to occur. If a nanoscale elastic structure exhibits a free energy barrier that can be surmounted by thermal fluctuations on physically relevant timescales, snap-through transitions would occur spontaneously with an associated rate constant, rendering the system metastable. In this work, we design a hydrogen-terminated graphene nanoribbon (GNR) system subject to adjustable geometric constraints that mimic macroscopic snap-through configurations. This nanoscale setup gives rise to a complex free energy landscape with tunable metastability, thereby enabling thermally activated snap-through transitions to occur on controllable timescales. The resulting system can serve as a prototypical platform for exploring rare event dynamics in thermalized elastic nanostructures.

While early studies reported various effects of thermal fluctuations on the morphology of two-dimensional (2D) materials such as freestanding graphene~\cite{fasolino2007intrinsic,los2009scaling,neek2014thermal,havsik2018quantum}, only recently have spontaneous transitions with signatures of thermal activation attracted increasing attention \cite{granato2023dynamics,hanakata2024vibrations,granato2025thermal}. The behavior of such microscopic systems coupled to a thermal environment often entails a level of complexity that merits more nuanced scrutiny~\cite{jarzynski2017stochastic}. In particular, an accurate description of energetics and kinetics for activated morphological transitions typically relies on capturing the detailed free energy landscape and transition pathways~\cite{peters2017reaction,bussi2020using}. For the geometrically constrained GNR system under consideration, enhanced sampling methods, such as metadynamics~\cite{barducci2008well}, are essential for resolving the GNR's complex free energy landscape. We propose that rare event approaches, specifically the combination of enhanced sampling techniques and generalized transition state theory~\cite{schenter2003generalized}, are well suited for the theoretical and computational handling of thermally activated transitions in microscopic elastic systems, even without any prior assumption about system energetics.

Enhanced sampling methods, which allow free energy calculations~\cite{frenkel2023understanding,bussi2020using} and accelerated molecular dynamics of rare events~\cite{voter1997method,voter1997hyperdynamics,tiwary2013metadynamics,ray2023kinetics}, have found widespread applications in chemical, biological, and material systems. Understanding the behavior of metastable nanomechanical systems coupled to thermal environments poses computational challenges analogous to those encountered in previous applications. One would thus expect enhanced sampling to be a natural tool for investigating metastable nanomechanical systems, yet its use in such scenarios remains an open frontier. Here we show that enhanced sampling methods can be applied to the metastable nanomechanical system of interest to derive a well-defined Landau free energy profile, which in turn enables the use of generalized transition state theory to describe and predict the transition kinetics. The accuracy and robustness of this combined approach are demonstrated through analysis of the GNR's behavior under a tunable free energy landscape. This methodology, long established in statistical mechanics and physical chemistry, can thus be extended to emerging activated processes in microscopic elastic systems, facilitating the design of nanoscale thermal switches and actuators.

\section*{Results}
\subsection*{\label{sec:A}Thermally Activated Snap-through Transitions}

A GNR can serve as a nanoscale counterpart of elastic strips, which have been increasingly adopted as a canonical system for analyzing the dynamics and mechanisms of snap-through transitions in bistable structures \cite{gomez2017critical,gomez2018pull,pandey2014dynamics,sano2017slip,sano2018snap,radisson2023elastic,radisson2023dynamic,wang2024transient}. Here we consider a zigzag GNR with a length of 15.23~nm and N = 16 zigzag chains, whose edges are hydrogen-terminated to eliminate intrinsic mechanical instabilities \cite{bets2009spontaneous,nikiforov2014formation}. Given the large length and time scales involved in the processes of interest, molecular dynamics (MD) simulations with empirical interatomic potentials constitute the only viable and effective computational approach that retains atomistic resolution~\cite{penev2021theoretical,dumitrica2010trends}.  The second generation reactive empirical bond order (REBO-II) potential \cite{brenner2002second} is adopted to describe our hydrogen-terminated GNR system.

\begin{figure}[t]
    \centering
    \includegraphics[width=0.48\textwidth]{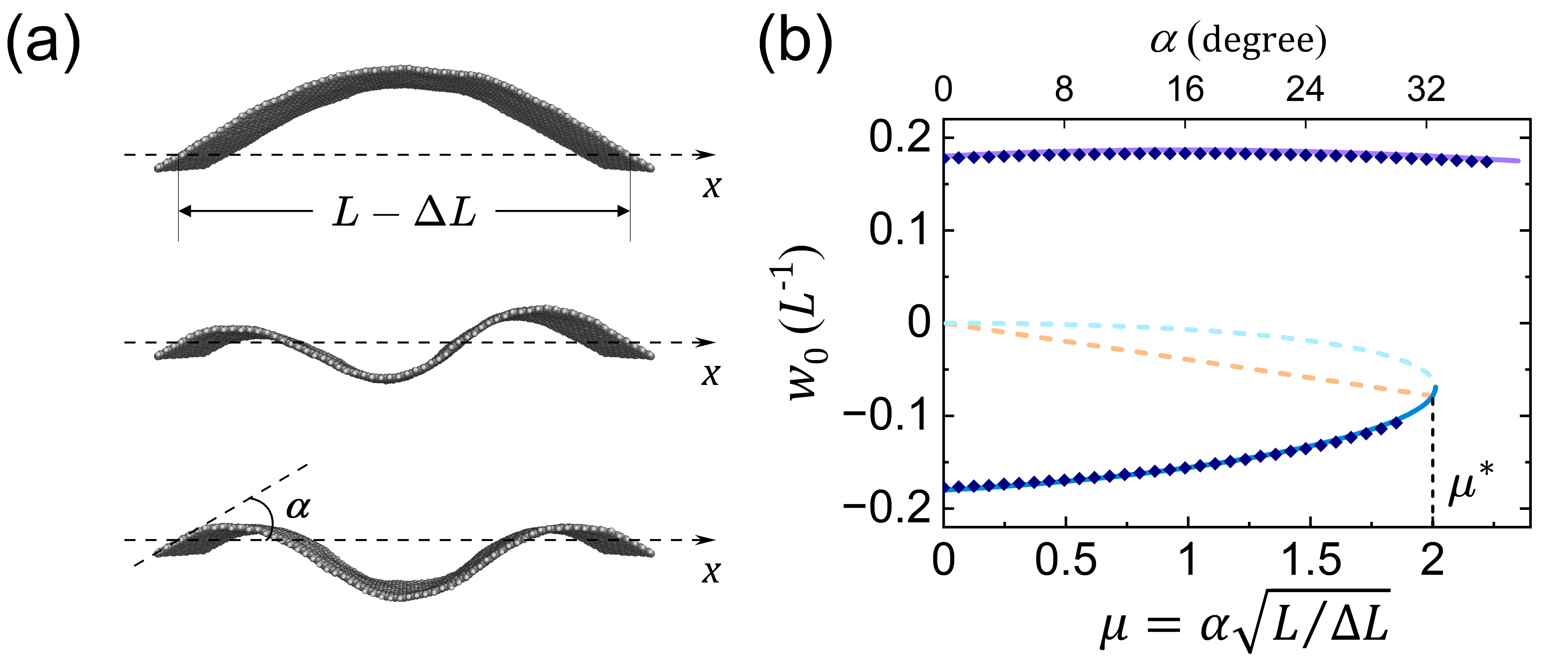}
    \caption{(a) Representative configurations of the GNR corresponding to the natural state (top), the asymmetric unstable equilibrium ($\mathcal{A}$ state, middle), and the inverted state (bottom). The rigidly constrained atoms at the two ends impose a fixed end-shortening $\Delta L$ and a symmetric tangent angle $\alpha$ at the boundaries. (b) The bifurcation diagram for the GNR system under consideration. Based on the equilibrium solutions of the Euler–beam equation, the midpoint displacement $w_0$ is plotted as functions of the geometric control parameter $\mu$, with solid lines indicating the natural state (violet upper branch) and the inverted state (blue lower branch), and the orange dashed line indicating the $\mathcal{A}$ state. Diamond markers represent the equilibrium averages of $w_0$ computed from MD simulations.}
    \label{fig:bifurcation}
\end{figure}

The GNR can be transformed into a bistable structure when subject to geometric constraint at the boundaries. To achieve this, the outermost seven rows of atoms at each end are rigidly held, leaving a flexible segment of length $L$ = 13.75~nm. A fixed end-shortening $\Delta L=0.08L$ is applied between the two ends, which are then rotated in opposite directions by a small adjustable angle $\alpha$, causing the GNR to adopt one of two mechanical stable states: a natural state and an inverted state, as shown in Fig.~\ref{fig:bifurcation}(a). The equilibrium configurations of these states can be analyzed with Euler–Bernoulli beam theory \cite{gomez2017critical,radisson2023elastic,wang2024transient} in the limit of small transverse displacement $w(x,t)$,
\begin{equation}
\rho bh\frac{\partial^2w}{\partial t^2}+B\frac{\partial^4w}{\partial x^4}+F\frac{\partial^2w}{\partial x^2}=0,
\label{eq:Euler}
\end{equation}
where $\rho$, $b$, and $h$ are the effective density, width, and thickness of the GNR, respectively; $B$ denotes the bending stiffness, and $F$ the applied compressive force. The geometric constraint enters as the boundary conditions and the end-shortening confinement condition \cite{gomez2017critical}. The time-independent solutions of Eq.~(\ref{eq:Euler}) capture the equilibrium behavior of the system, which can be summarized in a bifurcation diagram in Fig.~\ref{fig:bifurcation}(b). The midpoint displacement $w_0$ of the GNR, weighted by $L$, is plotted as a function of the dimensionless control parameter $\mu=\alpha\sqrt{L/\Delta L}$. As $\mu$ increases, the inverted state loses its stability at $\mu^*= 2$, where it intersects an asymmetric unstable equilibrium branch (denoted as $\mathcal{A}$ state) in a subcritical pitchfork bifurcation. In a macroscopic system, $\mu^*$ corresponds to the onset of snap-through towards the natural state. As the inverted state further approaches $\mu^*_2 \approx 2.012$, it undergoes a saddle-node bifurcation and vanishes. Meanwhile, the natural state persists stably over the full range of $\mu$. Note that the equilibrium configurations are entirely controlled by the geometric parameter $\mu$ and are independent of the material parameters of the system~\cite{gomez2018ghosts}.

To compare the behavior of the GNR in MD simulations with the description of the Euler-beam equation, we monitor $w_0$ of the GNR at different values of $\mu$, by setting $\alpha$ to successive integer degrees. For each $\alpha$, the GNR is fully thermalized in the natural and the inverted states separately, and a 5 ns trajectory is initiated from each state to calculate the time average of $w_0$. A Langevin thermostat is used to provide the thermalizing environment at 300~K. The time-averaged $w_0$ from our MD simulations is found to be in good agreement with the Euler–beam equation, as shown in Fig.~\ref{fig:bifurcation}(b). Naturally, the GNR configuration does not remain stationary at the mechanical equilibrium position of the natural or inverted state, but instead fluctuates around it, thereby defining a thermodynamic equilibrium state. When $\alpha$ exceeds $30^{\circ}$ ($\mu \approx 1.851$), the GNR cannot persist in the vicinity of the inverted state throughout a 5~ns trajectory: it snaps to the natural state under the influences of thermal fluctuations that are not captured by Eq.~(\ref{eq:Euler}). This indicates that the thermal stability of the inverted state has already been compromised, even though the system is well below $\mu^*$, the threshold of mechanical instability.

\begin{figure}[h]
    \centering
    \includegraphics[width=0.48\textwidth]{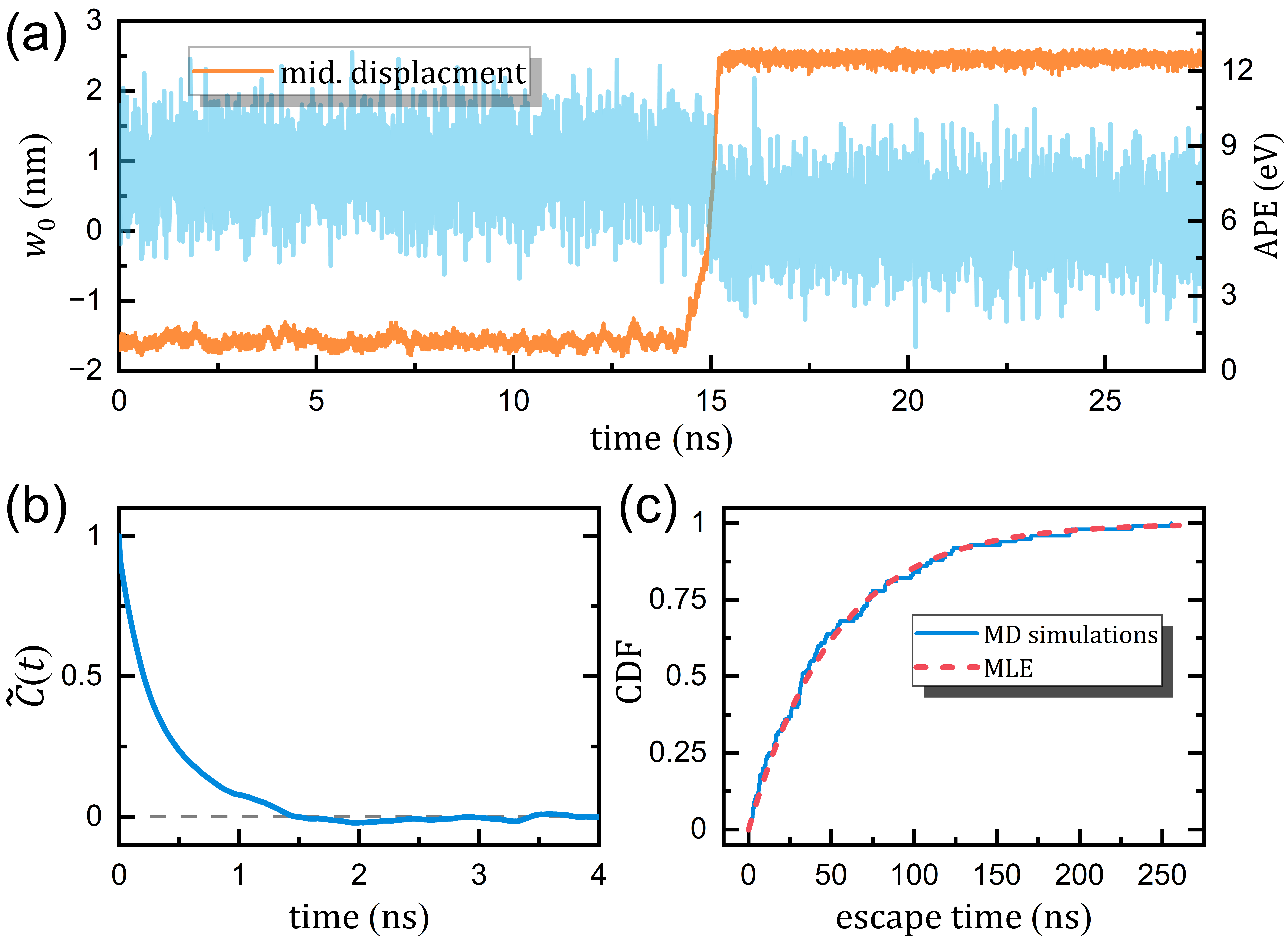}
    \caption{(a) Midpoint displacement $w_0$ of the GNR (left y-axis, orange) and the corresponding atomistic potential energy (APE) (right y-axis, light blue) in a representative trajectory of the thermally activated snap-through transition. The zero of APE has been shifted for clarity. (b) Normalized time-correlation function of $w_0$ within the local equilibrium of the inverted state. (c) Cumulative distribution function of the escape times from the inverted state (blue solid line), compared with the maximum likelihood estimation fit (red dashed line).}
    \label{fig:timescales}
\end{figure}

Accordingly, we systematically investigate thermally activated snap-through transitions by tracking the snap-through events in 100 trajectories with different initial conditions sampled from an equilibrium ensemble of the inverted state at $\alpha = 29^{\circ}$ ($\mu \approx 1.789$) and $T$ = 300 K. A representative transition trajectory is shown in Fig.~\ref{fig:timescales}(a). Prior to the snap-through event at $\approx$ 15 ns, the system stays in the local equilibrium of the inverted state, exhibiting small fluctuations of $w_0$. To characterize the relaxation of the system within the inverted state, we calculate the normalized time-correlation function of $w_0$, defined as
\begin{equation}
\widetilde{C}(t)=\frac{\langle\delta w_0(t)\delta w_0(0)\rangle}{\langle\delta w_0(0)\delta w_0(0)\rangle},
\label{eq:TCF}
\end{equation}
where $\delta w_0(t)=w_0(t)-\langle w_0 \rangle$, and $\langle \cdot \rangle$ denotes the ensemble average in the inverted state. The profile of $\widetilde{C}(t)$ is presented in Fig. \ref{fig:timescales}(b). The time integral of $\widetilde{C}(t)$ yields the dominant local relaxation time scale $\tau_s \approx$ 0.32~ns. In addition, we analyze the distribution of escape times in the 100 trajectories. The mean escape time is estimated using maximum likelihood estimation (MLE), yielding a value of $\tau_e = 51.70$ ns, which is two orders of magnitude greater than the local relaxation time $\tau_s$. The cumulative distribution function (CDF) of the escape times is shown in Fig. \ref{fig:timescales}(c), along with the fitted exponential distribution obtained from MLE. A Kolmogorov–Smirnov test produces a $p$-value of 0.957, confirming that the escape times follow an exponential distribution. The separation of time scales ($\tau_e \gg \tau_s$) \cite{hanggi1990reaction,peters2017reaction,chandler1978statistical} and the exponential distribution of escape times each demonstrate that the snap-through transitions in the GNR constitute a well-defined Poisson process.

\subsection*{Free Energy Landscape and Transition Pathways}

As shown in Fig. \ref{fig:timescales}(a), even while remaining within the local equilibrium states, the system exhibits substantial fluctuations in the atomistic potential energy, which obscures a clear description of the transition energetics. Moreover, the system behavior is subject to entropic effects, such as the configurational entropy~\cite{palazzesi2017conformational}. Therefore, instead of relying on elastic energy, a free-energy-based approach becomes necessary to capture the thermally activated snap-through at the nanoscale. Here, to determine the free energy profile of the system, we utilize well-tempered metadynamics (WT-MetaD) \cite{barducci2008well}, an enhanced sampling method that has been extensively used in complex molecular systems to accelerate rare events and accurately calculate free energy \cite{bussi2020using,dama2014well}.

\begin{figure*}[t]
    \centering
    \begin{minipage}[b]{0.4\textwidth}
    \includegraphics[width=\textwidth]{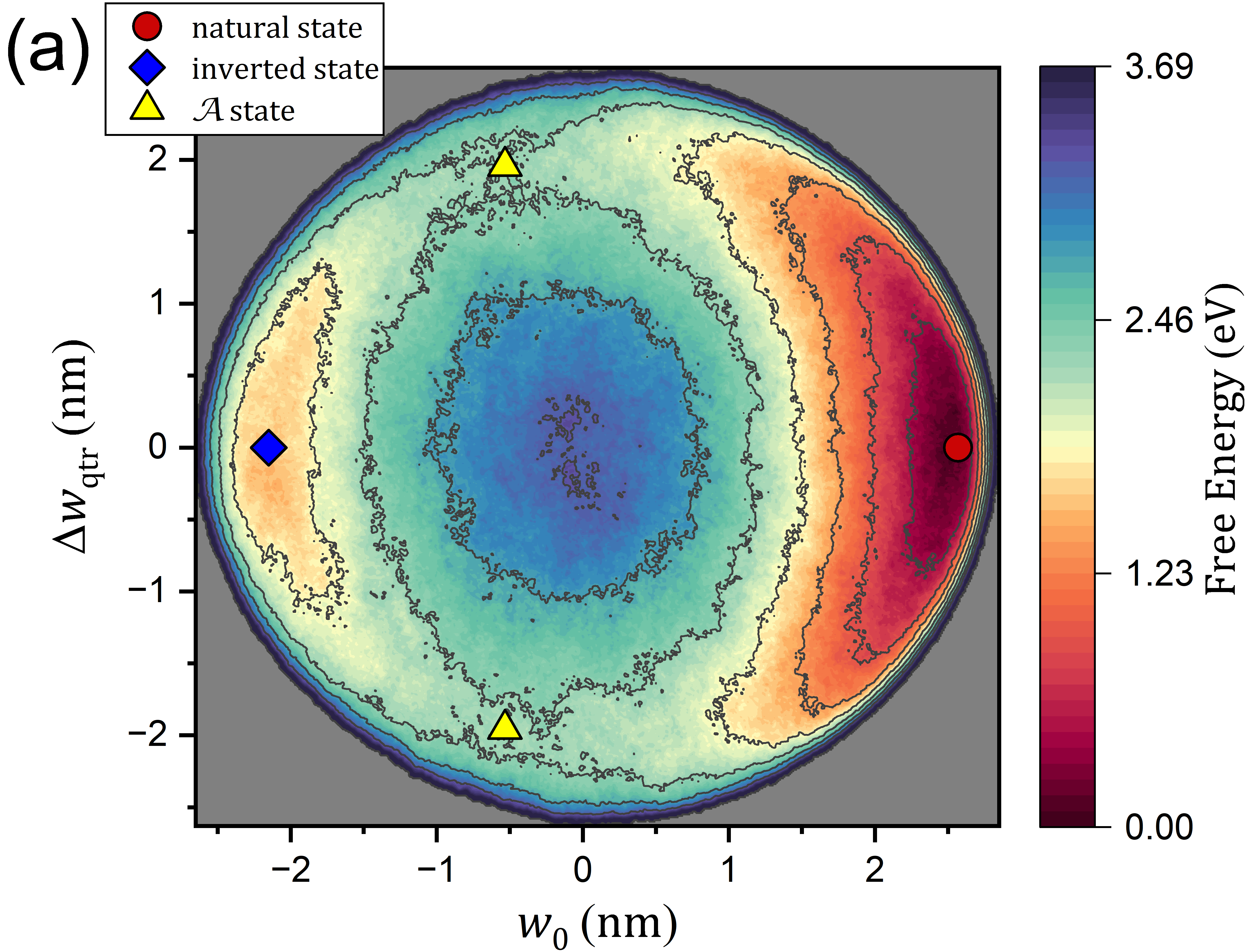}
    \end{minipage}
    \begin{minipage}[b]{0.4\textwidth}
    \includegraphics[width=\textwidth]{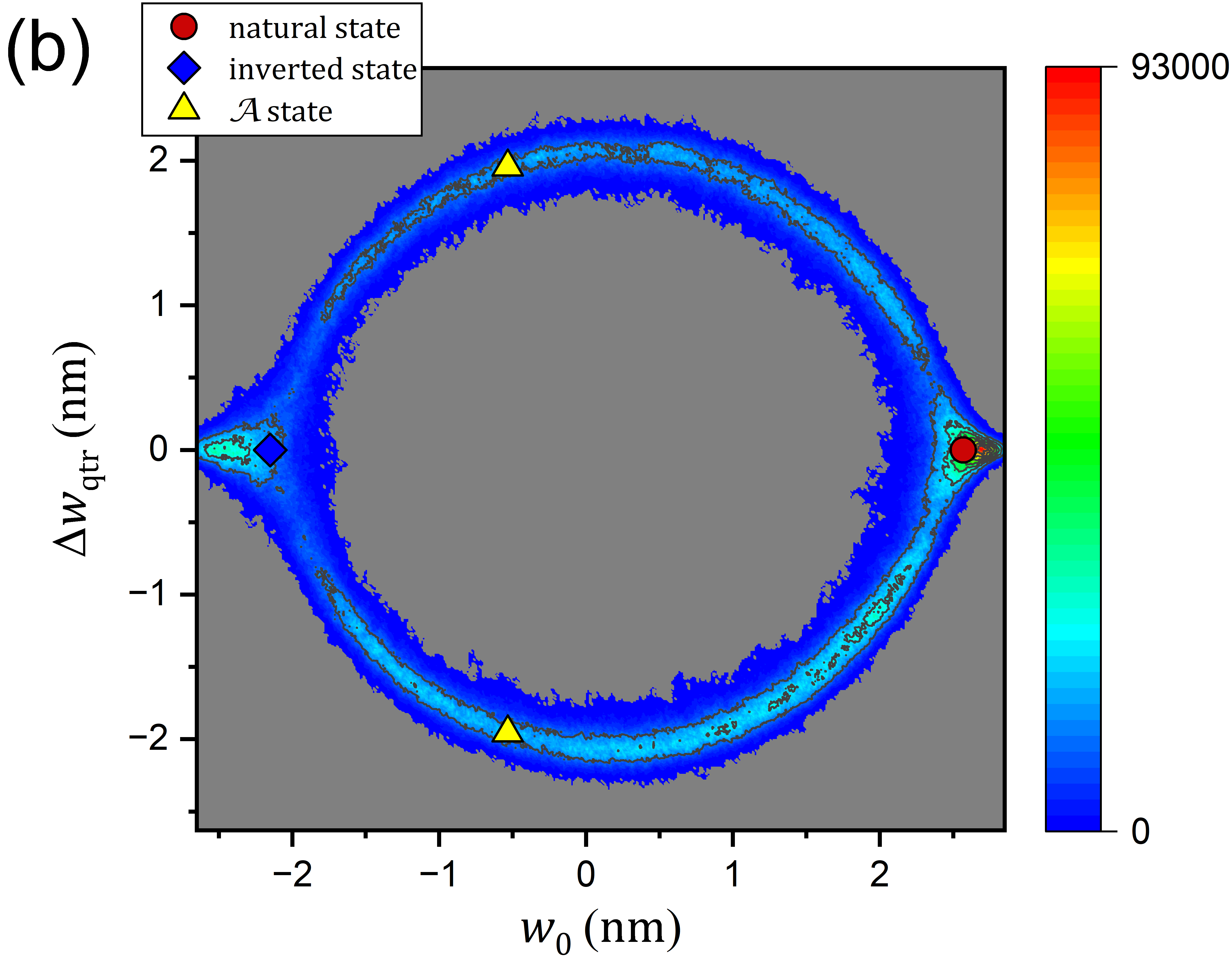}
    \end{minipage}
    \caption
    {(a) Free energy surface (FES) computed from 2D WT-MetaD in the ($w_0$, $\Delta w_{\text{qrt}}$) space, and (b) unreweighted histogram from a representative 1D WT-MetaD simulation biasing only $w_0$, as projected into the ($w_0$, $\Delta w_{\text{qrt}}$) space. The collective variable $\Delta w_{\text{qrt}}$ quantifies the degree of asymmetry of the GNR. The locations of the natural state (red circle), inverted state (blue diamond), and the $\mathcal{A}$ state (yellow triangle) are computed from the Euler-beam equation. The minimum free energy paths (MFEPs) on the FES in (a) are consistent with the transition pathways revealed by 1D WT-MetaD in (b). The results shown correspond to $T$ = 300 K and $\alpha = 16^\circ$ ($\mu \approx 0.9873$), exhibiting qualitative features common across the computed range $0 \leq \mu \lesssim 1.789$.}
    \label{fig:landscape}
\end{figure*}

2D WT-MetaD simulations are performed by biasing two carefully chosen collective variables: Since $w_0$ characterizes the progression of the snap-through transitions, it serves as the natural order parameter; in addition, motivated by the important role of symmetry breaking in macroscopic elastic strips \cite{radisson2023elastic,wang2024transient}, we propose a measure of asymmetry that is readily amenable to biased sampling in WT-MetaD, defined as the difference in transverse displacement between the quarter-length positions measured from the two ends of the GNR: $\Delta w_{\text{qrt}}=w(L/4) - w(-L/4)$. The free energy surface (FES) for the activated snap-through transitions is calculated in the state space of ($w_0$, $\Delta w_{\text{qrt}}$), as shown in Fig.~\ref{fig:landscape}(a). Two basins corresponding to the inverted state and the natural state are identified, with local minima aligned with $\Delta w_{\text{qrt}} = 0$, indicative of their symmetric nature. The transition pathways between the two states are characterized by the minimum free energy paths (MFEPs) connecting the basins, along which the system transitions between the two symmetric states via highly asymmetric intermediate states. To compare with elastic continuum theory, the stable mechanical equilibrium positions from the Euler-beam equation are plotted on the FES, coinciding with the metastable minima. Furthermore, the asymmetric unstable equilibria ($\mathcal{A}$ state) coincide with a pair of first-order saddle points, indicating their role as the transition state in thermal activations.

One can determine the one-dimensional (1D) Landau free energy profile along the order parameter $w_0$ by integrating the Boltzmann weight over $\Delta w_{\text{qrt}}$~\cite{peters2008path}, or equivalently, by WT-MetaD simulations biasing only $w_0$, as long as the 1D WT-MetaD trajectories faithfully track the MFEPs on the 2D free energy landscape (as confirmed by Fig.~\ref{fig:landscape}). Either way, it is essential to capture the pair of asymmetric pathways in order to correctly characterize the transition energetics.

\begin{figure}[h]
    \centering
    \includegraphics[width=0.48\textwidth]{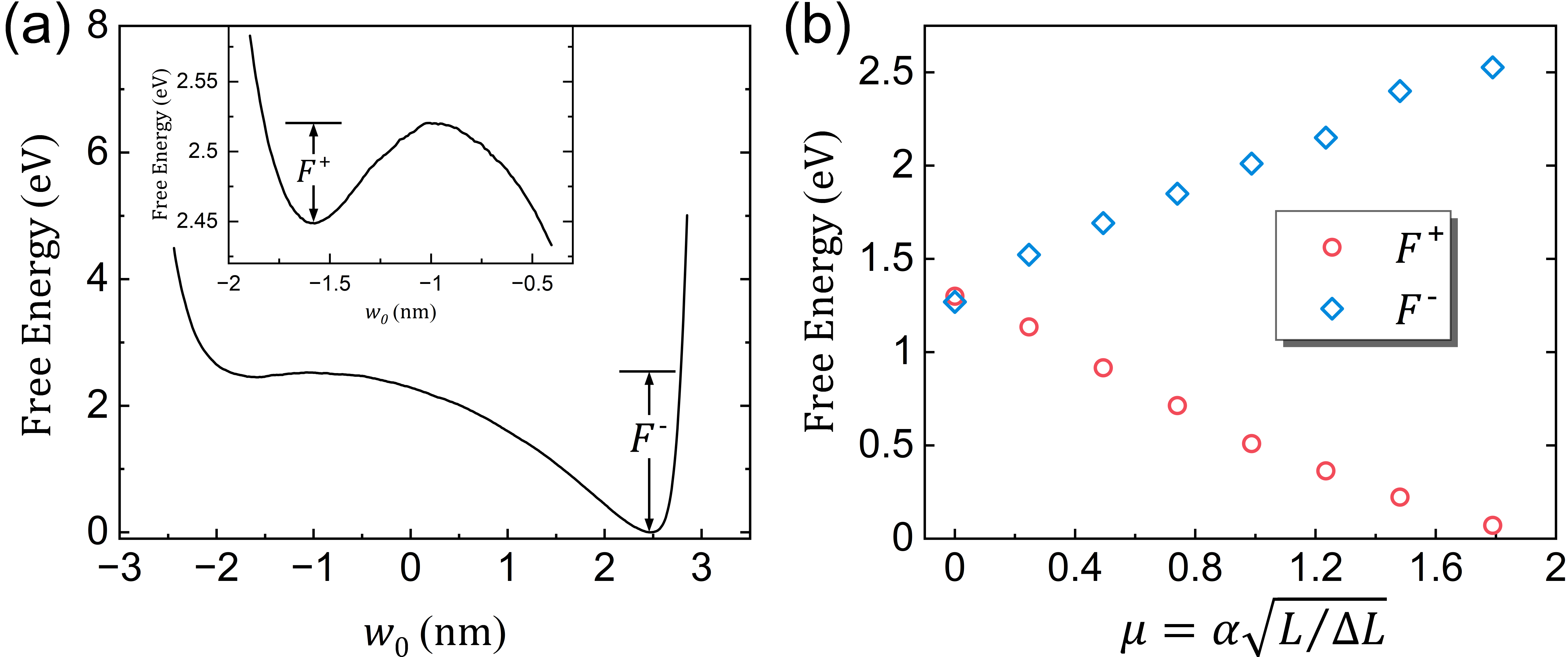}
    \caption{(a) Free energy profile computed from WT-MetaD (five 500 ns production runs) along the order parameter $w_0$ for the GNR constrained at $\alpha = 29^{\circ}$ ($\mu \approx 1.789$) under temperature $T$ = 300 K. Inset: Magnified view of the free energy profile highlighting the forward transition barrier $F^+$. (b) Forward and backward transition barriers, $F^+$ and $F^-$, as functions of the dimensionless geometric control parameter $\mu$.}
    \label{fig:Barriers}
\end{figure}

Corresponding to the transitions recorded in the unbiased simulations at $\alpha = 29^{\circ}$ under $T$ = 300 K, the free energy profile $F(w_0)$ calculated from 1D WT-MetaD is presented in Fig. \ref{fig:Barriers}(a), exhibiting two local minima corresponding to the inverted state and the natural state. The forward transition from the inverted to the natural state is associated with a free energy barrier of $F^+$ = 71.823~meV, while the backward transition corresponds to a much higher barrier of $F^-$ = 2526.776~meV. The substantial height of $F^-$ suggests that the reverse transition is kinetically suppressed, as will be addressed in our later analysis. Since the energetics of the system can be controlled by the geometric parameter $\mu$, we have further computed the free energy barriers from WT-MetaD at various values of $\mu$, as shown in Fig. \ref{fig:Barriers}(b). When $\mu = 0$, the forward and backward transition barriers are equal, implying no energetic preference between the two states. As $\mu$ increases, the barriers become increasingly unequal, showing that the metastability of the GNR system is tunable.

\subsection*{Generalized Transition State Theory in Terms of the Landau Free Energy}

Reaction rate theory has been widely employed to describe the kinetics of activated rate processes~\cite{hanggi1990reaction,peters2017reaction}, prominently including Kramers' theory~\cite{mel1991kramers} and transition state theory (TST)~\cite{truhlar1996current,bao2017variational}. In particular, generalized transition state theory in terms of the potential of mean force~\cite{schenter2003generalized} (or Landau free energy, depending on the context~\cite{ciccotti2025foundations}) provides a rigorous framework for characterizing rare event transition rates when used in conjunction with free energy calculation methods~\cite{watney2006calculation}. One can adopt this combined approach to comprehend the kinetics of thermally activated snap-through transitions.

Provided that there exists a collective variable $\xi$ such that the transition state dividing surface $\xi(\mathbf{r}) = \xi^\ddagger$ separates the reactant region $\{ \mathbf{r} | \xi(\mathbf{r}) < \xi^\ddagger \}$ from the product region $\{ \mathbf{r} | \xi(\mathbf{r}) > \xi^\ddagger \}$ in the configuration space, the transition rate constant can be calculated by the reactive flux through this surface. For a system characterized by the 1D Landau free energy $F(\xi)$, the generalized TST rate constant takes the form
\begin{align}
    k_\mathrm{TST} &= \frac{\langle {Z_\xi}^\frac{1}{2} \rangle_{\xi=\xi^\ddagger}}{(2\pi\beta)^\frac{1}{2}}\frac{\mathrm{e}^{-\beta F(\xi^\ddagger)}}{\int_{\xi < \xi^\ddagger} d\xi\, \mathrm{e}^{-\beta F(\xi)}} \label{eq:rate_const1} \\ 
    &= \frac{\langle {Z_\xi}^\frac{1}{2} \rangle_{\xi=\xi^\ddagger}}{(2\pi\beta)^\frac{1}{2} {\Delta\xi}} \mathrm{e}^{-\beta \Delta F^\ddagger}. \label{eq:rate_const2}
\end{align}
where $\langle \cdot \rangle_{\xi=\xi^\ddagger}$ denotes the ensemble average sampled at the transition state and $\int_{\xi < \xi^\ddagger}$ denotes the integral over the reactant region. Here $\beta$ is the inverse temperature, $\Delta\xi$ is given by $\Delta\xi=\int_{\xi < \xi^\ddagger} d\xi$, $\Delta F^\ddagger$ is the free energy of activation, given by
\begin{equation}
\Delta F^\ddagger=F(\xi^\ddagger)+k_\mathrm{B}T\ln{({\Delta \xi}^{-1} \int_{\xi < \xi^\ddagger}d\xi\,\mathrm{e}^{-\beta F(\xi)})},
\label{eq:del_F}
\end{equation}
and $Z_\xi$ is the inverse effective mass associated with $\xi$, defined as
\begin{equation}
Z_\xi\equiv\sum_{i=1}^{3N}\frac{1}{M_i}\left(\frac{\partial\xi}{\partial r_i}\right)^2,
\label{eq:inverse_mass}
\end{equation}
with $M_i$ being the atomic mass associated with the $i$th component of $\mathbf{r}$. For the GNR system under investigation, the order parameter $w_0$ serves the role of $\xi$. It should be noted that the free energy of activation $\Delta F^\ddagger$ is not identical to the transition barrier $F^+$ due to a statistical distribution of the reactant state~\cite{schenter2003generalized}.

\begin{figure}[t]
    \centering
    \includegraphics[width=0.48\textwidth]{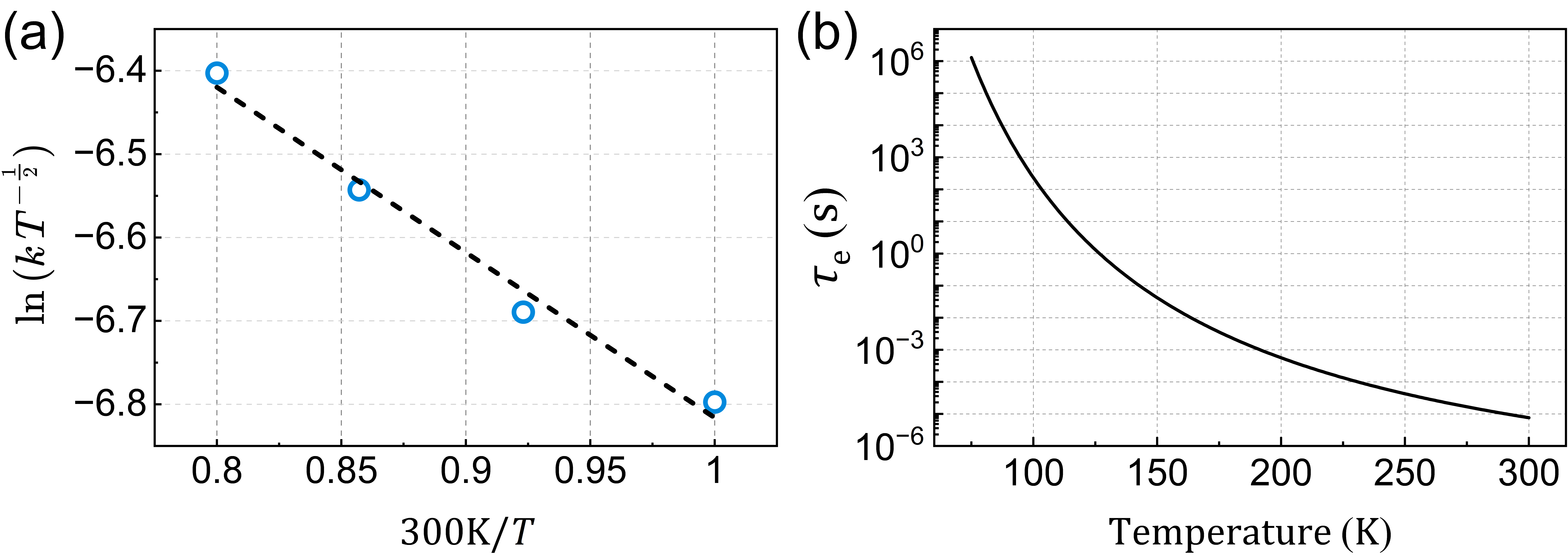}
    \caption{(a) For the GNR system constrained at $\alpha = 29^{\circ}$, $\ln{(k\,T^{-\frac{1}{2}})}$ is plotted versus $1/T$ to reveal the temperature dependence of the transition rate constant $k$ computed directly from unbiased simulations. Based on Eq.~(\ref{eq:rate_const2}), $\Delta F^\ddagger$ can be inferred from the slope of the linear fit. (b) Transition rate constant $k$ for the forward transition as a function of temperature, calculated using Eq.~(\ref{eq:rate_const1}) and WT-MetaD for the system constrained at $\alpha = 24^{\circ}$.}
    \label{fig:T_dependence}
\end{figure}

We justify the applicability of generalized TST to our system by examining the consistency between the free energy of activation $\Delta F^\ddagger$ inferred from Eq.~(\ref{eq:rate_const2}) based on the transition rates determined in unbiased simulations at different temperatures and the value of $\Delta F^\ddagger$ calculated directly from $F(w_0)$ given by WT-MetaD. To this end, when the system is constrained at $\alpha = 29^{\circ}$, we first calculate $\Delta F^\ddagger$ in the temperature range from 300~K to 375~K by Eq.~(\ref{eq:del_F}). The resultant $\Delta F^\ddagger$ varies from 48.817~meV to 49.775~meV, showing that it is approximately constant over the temperature range considered.

In parallel, unbiased MD simulations are conducted to record the escape times for transitions from the inverted state to the natural state at 300~K, 325~K, 350~K, and 375~K, respectively. At each temperature, the transition rate constant $k$ is obtained as the inverse of the mean escape time $\tau_e$ from 100 independent trajectories (see the Supporting Information). To analyze the temperature dependence of $k_\mathrm{TST}$, we recall that $w_0$ is computed as a linear combination of the midline C atom coordinates, leaving $\langle {Z_{w_0}}^\frac{1}{2} \rangle_{w_0={w_0}^\ddagger}$ constant, and the temperature dependence of $k_\mathrm{TST}$ lies in the prefactor $T^\frac{1}{2}$ and the Boltzmann factor of Eq.~(\ref{eq:rate_const2}). Motivated by this observation, we plot $\ln{(k\,T^{-\frac{1}{2}})}$ against $1/T$ in Fig.~\ref{fig:T_dependence}(a), where a linear trend indicates that $k$ follows the temperature dependence described by Eq.~(\ref{eq:rate_const2}). Moreover, the slope of the linear fit gives $\Delta F^\ddagger$ = 51.228~meV, which is in good agreement with the $\Delta F^\ddagger$ values previously calculated from WT-MetaD (average relative error $< 5\%$). These findings demonstrate that the temperature dependence of the rate constant $k$ is accurately captured by generalized TST, and corroborate the self-consistency of our computational results. In practice, the rate constant $k$ generally deviates from the TST prediction $k_\mathrm{TST}$ by a multiplicative factor, the transmission coefficient $\kappa$, which accounts for recrossing effects. However, the consistent temperature dependence exhibited by $k$ and $k_\mathrm{TST}$ here implies that, for our system, $\kappa$ remains approximately constant over at least a moderate range of temperature.

While generalized TST alone does not directly yield the absolute value of the rate constant $k$, Eq.~(\ref{eq:rate_const1}) still allows us to predict the kinetics of the GNR system under various conditions by comparing relative changes in $F(w_0)$ and temperature. To illustrate this, the geometric constraint is adjusted from $\alpha = 29^\circ$ to $24^\circ$, and the updated $F(w_0)$ is calculated with WT-MetaD at room temperature. Based on the difference of $F(w_0)$ in the two settings, Eq.~(\ref{eq:rate_const1}) predicts $k=66.68$~ms$^{-1}$ at $\alpha = 24^\circ$ and $T$ = 300 K. Meanwhile, we compute the rate constant by the accelerated MD of hyperdynamics~\cite{voter1997method,voter1997hyperdynamics} under the same conditions, obtaining $k=77.69 \pm 21.54$~ms$^{-1}$ (see the Supporting Information).  The prediction of Eq.~(\ref{eq:rate_const1}) falls within the 95\% confidence interval of the hyperdynamics computation result. Thus, the predictive performance of generalized TST is robust across variations in the free energy landscape that are realized by changing the geometric confinement.

Fig.~\ref{fig:T_dependence}(b) presents the evolution of the rate constant $k$ computed by Eq.~(\ref{eq:rate_const1}) based on the WT-MetaD free energy calculations at $\alpha = 24^\circ$ and various temperatures. A temperature variation of $\Delta T = 25$ K can result in up to orders-of-magnitude changes in $k$. In addition, the backward transition is associated with $\Delta F_\text{back}^\ddagger = 2.31$~eV at 300~K, corresponding to a mean escape time of $\sim 10^{23}$ years, which indicates that the natural state is effectively stable under this condition. This construction can be exploited to control the transition direction and is particularly useful in the application of a thermal switch or actuator. Our analysis highlights that temperature control, along with the adjustment of activation free energy through geometric confinements (as illustrated in Fig.~\ref{fig:T_dependence}(b)), can offer an effective strategy for tuning the transition rate, its temperature sensitivity, and the transition direction of the activated snap-through in the GNR system.

\section*{Discussion}

The effects of thermal fluctuations on 2D materials are ubiquitous, but thermally activated morphological transitions in these systems are uncommon under ambient conditions. This is because realizing a well-defined activated process typically requires an activation free energy of a certain order of magnitude to produce observable transitions on accessible timescales. The GNR system discussed in this work exhibits a free energy landscape that is tunable through geometric confinement, thereby providing a prototype for realizing and regulating thermally activated morphological transitions in 2D materials. Moreover, recent years have seen a surge of interest in bistable elastic structures that display snap-through behavior~\cite{gomez2017critical,gomez2018pull,pandey2014dynamics,sano2017slip,sano2018snap,radisson2023elastic,radisson2023dynamic,wang2024transient}. When scaled down to the nanoscale, such structures may offer metastable free energy landscapes controlled by external constraints, suggesting a promising route for engineering thermally activated mechanical transitions at the nanoscale.

In this work, we introduce enhanced sampling methods, including WT-MetaD and hyperdynamics into the study of nanoscale elastic systems exhibiting metastability. WT-MetaD enables us to reconstruct the complex free energy landscape and to reveal a pair of asymmetric transition pathways of the geometrically constrained GNR system. The resultant Landau free energy profile is compatible with the application of generalized transition state theory, which allows us to accurately capture the kinetics of the system over a range of temperature and geometric constraints. This methodology can be readily generalized to other bistable elastic nanostructures coupled to a thermalizing environment. For more complex structures, even in the absence of prior knowledge about the optimal collective variables, data-driven collective variable construction methods can be integrated with WT-MetaD to recover nontrivial free energy landscapes~\cite{wang2020machine,wang2019past,mehdi2024enhanced,zhu2025enhanced}. Our findings offer a theoretical framework for the study of elastic
metastability, advance the understanding of thermalized nanomechanical systems, and provide design strategies for tunable nanoscale thermal switches and thermal actuators.

\section*{Materials and Methods}

All simulations were performed with the LAMMPS \cite{plimpton1995fast} package. The atomic interactions of the studied system were described by the REBO-II potential \cite{brenner2002second}, which is one of the most commonly used empirical potentials for solid carbon and hydrocarbon systems and has been shown to reliably capture the elasticity \cite{lu2009elastic} and thermomechanical behaviors \cite{,gao2014thermomechanics} of graphene. The outermost seven rows of atoms at each end of the GNR were rigidly held to implement the geometric constraint, and the other atoms evolved according to the velocity-Verlet algorithm with a time step of 1.0 fs. A Langevin thermostat was used to provide the thermalizing environment at specified temperatures with a damping coefficient of 1.0 ps$^{-1}$ \cite{savin2022friction,li2019influence,savin2023critical}. The initial configurations of unbiased simulations were obtained from sampling the equilibrium ensemble of the inverted state or the natural state. WT-MetaD and hyperdynamics simulations were carried out with the COLVARS module \cite{fiorin2013using}. The WT-MetaD parameters were optimized when the system boundaries were constrained at each angle $\alpha$. The order parameter $w_0$ was computed as the transverse displacement of the center of mass of the C atoms in the longitudinal middle row of the GNR. The collective variable $\Delta w_{\text{qrt}}$ was defined as $\Delta w_{\text{qrt}} = w(L/4) - w(-L/4)$, where $w(L/4)$ and $w(-L/4)$ denote the average transverse displacements of the C atom rows located at the quarter-length positions measured from each end of the GNR. Further methodological and computational details can be found in the Supporting Information.

\begin{acknowledgments}
R.Z., Y.Z. and C.L. gratefully acknowledge support from the National Natural Science Foundation of China under Grant No. 21973046.
Y.W. gratefully acknowledges support from the startup funds provided by Case Western Reserve University.
\end{acknowledgments}

\bibliography{mybib}

\end{document}